\documentclass[twoside]{JHEP3}
\title{Holographic Meson Melting}
\author{Carlos Hoyos,${}^a$ Karl Landsteiner${}^b$ and Sergio Montero${}^b$\\
${}^a$Department of Physics,
University of Wales Swansea\\
~\,Swansea, SA2 8PP, UK\\
~\,E-mail: \email{C.H.Badajoz@swansea.ac.uk}\\
${}^b$Instituto de F\'{\i}sica Te\'orica, C-XVI Universidad Aut\'onoma de Madrid\\
~\,E-28049 Madrid, Spain\\
~\,E-mail: \email{Karl.Landsteiner@uam.es, Sergio.Montero@uam.es}}
\abstract{The plasma phase at high temperatures of a strongly coupled gauge theory can be holographically modelled by an AdS black hole. Matter in the fundamental representation and in the quenched approximation is introduced through embedding D7-branes in the AdS-Schwarzschild background. Low spin mesons correspond to the fluctuations of the D7-brane world volume. As is well known by now, there are two different kinds of embeddings, either reaching down to the black hole horizon or staying outside of it. In the latter case the fluctuations of the D7-brane world volume represent stable low spin mesons. In the plasma phase we do not expect mesons to be stable but to melt at sufficiently high temperature. We model the late stages of this meson melting by the quasinormal modes of D7-brane fluctuations for the embeddings that do reach down to the horizon. The inverse of the imaginary part of the quasinormal frequency gives the typical relaxation time back to equilibrium of the meson perturbation in the hot plasma. We briefly comment on the possible application of our model to quarkonium suppression.}
\preprint{IFT-UAM/CSIC-06-58\\ SWAT/06/484\\ \texttt{hep-th/0612169}}
\usepackage[english]{babel}
\usepackage{amssymb}
\usepackage{amsfonts}
\usepackage{amsmath}
\usepackage{graphicx}
\usepackage{epsfig}
\usepackage{psfrag}
\usepackage{cite}
\def\erf#1{(\ref{#1})} 
\newcommand{\cO}{{\cal O}}

  \newcommand{\bbR}{{\mathbb R}}

\newcommand{\bE}{{\mathbf E}}

  \newcommand{\bZ}{{\mathbf Z}}

\newcommand{\bg}{{\mathbf g}}  \newcommand{\by}{{\mathbf y}}

\newcommand{\be}{\begin{equation}}     \newcommand{\ee}{\end{equation}}
\newcommand{\bea}{\begin{eqnarray}}    \newcommand{\eea}{\end{eqnarray}}
\newcommand{\beann}{\begin{eqnarray*}} \newcommand{\eeann}{\end{eqnarray*}}
\newcommand{\bfig}{\begin{figure}}     \newcommand{\efig}{\end{figure}}
\newcommand{\ba}{\begin{array}}        \newcommand{\ea}{\end{array}}
\newcommand{\bcen}{\begin{center}}     \newcommand{\ecen}{\end{center}}
\newcommand{\btab}{\begin{tabular}}    \newcommand{\etab}{\end{tabular}}
\newcommand{\nn}{\nonumber}
\def\U{\operatorname{U}}         \def\SU{\operatorname{SU}}

\renewcommand{\Re}{\mathop{\rm Re}}   \renewcommand{\Im}{\mathop{\rm Im}}
\newcommand{\vev}[1]{\left\langle{#1}\right\rangle}
\newcommand{\then}{\Rightarrow}
\newcommand{\dd}{{\rm d}}
\newcommand{\e}{{\rm e}}
\newcommand{\adsfive}{$\mathop{\rm AdS_5\times{}S^5}$\,}
   \def\Ntwo{{\cal N}\!=\!2}   \def\Nfour{{\cal N}\!=\!4}

\newtheorem{Proposition}{Proposition}[section]

\newtheorem{Theorem}{Theorem}[section]
\newtheorem{Lemma}{Lemma}[section]
\newtheorem{Corrolary}{Corrolary}[section]

\newcommand{\bp}{\begin{Proposition}}	\newcommand{\ep}{\end{Proposition}}
\newcommand{\bt}{\begin{Theorem}}	\newcommand{\et}{\end{Theorem}}
\newcommand{\bl}{\begin{Lemma}}		\newcommand{\el}{\end{Lemma}}
\newcommand{\bc}{\begin{Corrolary}}	\newcommand{\ec}{\end{Corrolary}}
\begin{document}

\section{Introduction}  \label{sec:intro}
The AdS/CFT correspondence \cite{firstadscft} allows to model non-Abelian gauge theories in the strong coupling regime through a dual theory of gravity. The best understood example by far is the case of $\Nfour$ gauge theory and type IIB superstrings on the \adsfive background. The $\Nfour$ theory is rather special as it possesses exact conformal invariance and therefore stays in a Coulomb phase even at infinite strong coupling. The conformal symmetry corresponds to the isometry group of the five-dimensional AdS space. If one considers the gauge theory at finite temperature, thereby breaking the conformal symmetry, it immediately goes over into a deconfinement phase. The holographic dual of this deconfinement phase is given by the anti-de Sitter Schwarzschild black hole, more precisely by the AdS black hole with flat horizon geometry \cite{Witten:1998zw}. This allows to model a non-Abelian gluon plasma (of infinite extent) at strong coupling.

Recently this has become of increased interest since it turned out that the state of matter created in heavy-ion collisions at RHIC is most likely described by a strongly coupled quark-gluon plasma. Since more traditional methods such as finite temperature field theory or lattice QCD have a hard time to model the dynamics of the quark-gluon plasma at strong coupling, the AdS/CFT correspondence is a promising alternative to do this. In fact, several important plasma parameters, most notably the shear viscosity \cite{viscosity} and the energy loss rate of a heavy quark \cite{dragforce} or the jet quenching parameter using light-like Wilson loops \cite{jetquenching}\footnote{Although there is some controversy about the correct prescription, see \cite{Argyres:2006yz}.} have been calculated with stringy AdS-methods and in general turned out to be in surprisingly good agreement with experimental data \cite{Muller:2006ee}. It is therefore of highest importance to study the physics of the strongly coupled quark-gluon plasma as modelled by AdS black hole physics and to extend the range of physical processes that can be addressed in this way.

The spectrum of the $\Nfour$ theory contains only particles in the adjoint representation of the gauge group. This limitation can be overcome by giving the strings a place to end on, i.e. introducing D-branes. The endpoints of the string represent particles in the fundamental representation of the gauge group (``quarks"). Fundamental matter is modelled in this way through the embedding of D7-branes in \adsfive. So far, this has been achieved only in a sort of quenched approximation where the number of flavours $N_f$ is much less than the number of colours $N_c$.{}\footnote{See however \cite{Casero:2006pt} for recent progress beyond the quenched approximation.} In the AdS context it means that the D7-branes are introduced as probe branes in the gravity background, ignoring their backreaction onto the geometry \cite{Karch:2002sh}. A recent review on adding fundamental matter to the gauge/gravity correspondence can be found in \cite{Ramallo:2006et}.

Probe branes embedded in black hole geometries have already been studied some time ago in \cite{Frolov:1998td}, where it was found that there are two topologically distinct classes of embeddings. The first class, named Minkowski embedding, stays everywhere outside the black hole, whereas the second class, the black hole embedding, reaches down to the horizon such that the induced geometry on the brane itself is a black hole.

D7-brane embeddings in AdS black hole geometries have first been considered in \cite{Babington:2003vm}. Holography relates the asymptotic behaviour at the conformal boundary of the D7-brane to the quark mass $m_q$ and the quark bilinear condensate $\vev{q\bar q\,}$. The two types of embeddings give rise to a first order phase transition where the value of the quark condensate jumps by a finite amount. This phase transition has been studied in great detail in \cite{Kirsch:2004km, Apreda:2005yz, Albash:2006ew, Karch:2006bv, Mateos:2006nu}. This fundamental phase transition has been generalised to other types of backgrounds and D6- or D8-brane embeddings where there is a chiral $\U(1)$ or even a non-Abelian chiral symmetry, and the quark condensate is an order parameter for chiral symmetry breaking \cite{Kruczenski:2003uq, Sakai:2004cn}.

The model we consider in this paper, IIB sugra on \adsfive with \hbox{D7-branes} partially wrapped on the ${\rm S}^5$, is dual to $\Nfour$ $\SU(N_c)$ gauge theory at finite temperature $T$ in the large $N_c$ limit with one (or few) $\Ntwo$ hypermultiplets of mass $m_q$ in the fundamental representation. Because of the conformal symmetry we can either set $T$ or $m_q$ to one. Varying the mass of the quarks is therefore equivalent in this model to varying the temperature, i.e. lowering the mass is the same as rising the temperature. High values of the mass $m_q$ correspond to Minkowski embeddings and low values to black hole embeddings. Therefore we can also think of the two kind of embeddings as being at low temperature (Minkowski embedding) or high temperature (black hole embedding).

Quark anti-quark bound states give rise to mesons. In the holographic dual high spin mesons are modelled by spinning open strings ending on the probe branes. The decay of high spin mesons has been studied in \cite{mesonmeltinHighspin}. We will be concerned with the decay of low spin mesons in this paper. Low spin mesons are represented by the small fluctuations around the equilibrium configuration of the D-brane world volume. The spectrum of the D-brane fluctuations is very different for the two different classes of embeddings. In the case of the Minkowski embeddings there is a discrete spectrum of modes with eigenvalues that can be identified with the meson masses \cite{Karch:2002sh, Kruczenski:2003be}. Technically one imposes Dirichlet or Neumann boundary conditions for the fluctuations at the endpoints of the embedded brane. A wave travelling along the brane will then get reflected at the endpoint and
this gives rise to a discrete set of eigenmodes and eigenfrequencies. If the brane embedding is however such that it touches the horizon, a wave travelling down the brane will eventually reach the horizon, fall through it and not come back again (see figure \ref{fig:bcs}). In this case one has to impose purely infalling boundary conditions at the position of the horizon. These boundary conditions give rise to eigenmodes with complex frequencies, the so-called \textit{quasinormal modes.}

Quasinormal modes are damped fluctuations in black hole backgrounds and traditionally have been studied in the context of black holes in asymptotically flat space (see \cite{quasinormal_review} for a review). In the context of the AdS/CFT correspondence they have been first studied in \cite{horowitz_hubeny}. The imaginary part of the quasinormal frequency represents a decay constant. The dual field theory interpretation is that the inverse of the imaginary part gives the relaxation time back to equilibrium under a small perturbation of the quark-gluon plasma. Our interpretation of the quasinormal modes on the branes is as follows: the quasinormal modes represent the late stages of the melting process of a meson inserted in a plasma at high temperature. The imaginary part of the frequency
\begin{equation}
\omega^\mathrm{quasinormal}_n = \Omega_n - i\,\frac{\Gamma_n}{2} ~,
\end{equation}
gives the decay constant of the collective mode corresponding to the quasinormal mode in the plasma.\footnote{A holographic interpretation of the quasinormal modes in the case of the planar AdS black hole as poles in retarded Greens functions has been given in \cite{Kovtun:2005ev}. It also has been pointed out there that these poles can not be automatically given a quasiparticle interpretation. Our case is slightly different in that we are considering an $\Ntwo$ theory with an additional scale (the quark mass). Nevertheless we do not want to interpret the quasinormal modes as quasiparticles, solely as the collective modes representing the late time stages of the decay of a (meson)-perturbation inserted in the plasma.} In other words: mesons built up of sufficiently light quarks (or equivalently at sufficiently high temperature) inserted in the plasma will melt just as an icecube melts in hot water. At a late stage the typical timescales of this melting process is given by the inverse of the imaginary part of the quasinormal frequencies. We could start for example with a Minkowski type D7-brane embedding in an excited state with a 
normal mode fluctuation on it representing a stable meson. If we now increase the temperature slowly the D7-brane with the normal mode on it will eventually enter the unstable regime and undergo the phase transition to the black-hole embedding. Since the fluctuation is considered to be a small perturbation we can assume that the phase transition is basically unchanged from the one that takes place for the groundstate of the D7-brane. After the phase transition the additional energy present due to the meson perturbation will eventually drop into black hole and this decay process is goverened by the quasinormal modes. In this way the quasinormal modes are related to the melting of the meson at high temperature.

\FIGURE{\centering
\includegraphics[scale=0.9]{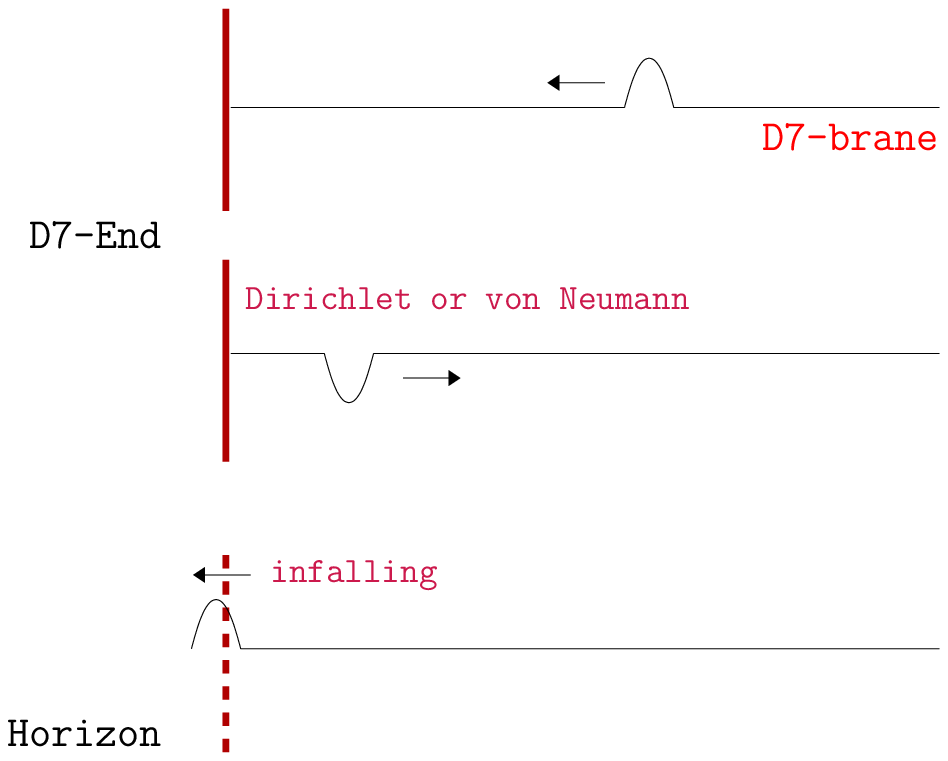}
\caption{\label{fig:bcs}Waves travelling down a D7-brane towards the interior of an AdS black hole. For Minkowski embeddings the wave meets the end of the brane and is reflected there. This gives rise to normal modes and frequencies that are the holographic duals of low spin mesons and their masses. In the case of the black hole embeddings the wave travels down the brane and through the horizon and simply never comes back. The fluctuation has to obey infalling boundary conditions giving rise to quasinormal modes!}}

The paper is organised as follows: in section \ref{sec:embeddings} we review the physics of the D7-brane embeddings in the AdS black hole geometry. It also serves to set up our notation and conventions.

In section \ref{sec:quasinormal} we first compute the quasinormal modes for the trivial embedding. This is the embedding
that corresponds to massless quarks in the dual field theory. The differential equation that the fluctuations obey is a Heun equation. This type of equation has appeared already several times before in the study of quasinormal modes. We adopt the method of continued fractions as pioneered by \cite{Leaver} and applied in the AdS context in \cite{Starinets:2002br}. The method converges very fast and allows a quick calculation of the first low-lying quasinormal modes. We present the results for the first ten quasinormal frequencies. Then we go over to brane embeddings that correspond to massive quarks. The branes are bent for these embeddings and the embedding itself can be calculated only through numerical integration of a highly non-linear differential equation. This makes the calculation of the quasinormal modes around these embeddings much more difficult. We employ two strategies. First we approximate the numerical embedding by an ansatz that is introduced in the differential equation of the fluctuations. We solve the equations at the boundary and at the horizon using series expansions with the right asymptotics. We use the horizon series to give initial values close to the horizon and integrate numerically towards the boundary. We then match the numerical solutions with the boundary series at a point close to the boundary and demand that the resulting solution is smooth. This fixes the quasinormal frequency. We were able to find the three lowest quasinormal modes for a number of brane embeddings up to masses that actually lie already above the critical mass, where the phase transition to the Minkowski embeddings occurs.

We also employed a second method to find the quasinormal modes. This second method consists in converting the differential equation into a finite difference equation. In this approach one imposes the correct boundary conditions at the endpoints and uses an ansatz solution that is fed into the finite difference equation. An improvement to the ansatz solution $y$ can be computed by demanding that $y+\Delta y$ fulfils the finite difference equations to first order, in an expansion in the correction $\Delta y$. This gives a new ansatz solution and therefore it amounts to a recursive algorithm that eventually relaxes to the correct quasinormal mode and frequency within a given accuracy goal. We describe this relaxation method in more detail in the appendix. In praxis, it turned out that the relaxation method worked well only for the lowest quasinormal mode. The agreement of the quasinormal frequencies as computed with the relaxation method or with the midpoint shooting algorithm is however excellent and typically of the order of
\begin{equation}
\frac{\left|\omega^{\rm shoot} - \omega^{\rm relax}\right|}{\left|\omega^{\rm shoot}+\omega^{\rm relax} \right|} \approx 10^{-4} ~.
\end{equation}
We are therefore confident that the quasinormal frequencies are accurate up to the quoted uncertainty. In this paper we will state only the results from the midpoint shooting method.

In section \ref{sec:mesonmasses} we briefly discuss the masses of the mesons that are represented by the fluctuations around the Minkowski embeddings. Finally, in section \ref{sec:conclude} we sum up our conclusions and give an outlook to possible further applications of our holographic picture of meson melting to the physics of the quark-gluon plasma.

\section{D7-brane embeddings}  \label{sec:embeddings}
The ten-dimensional geometry is given by the direct product of the five-dimensional flat AdS black hole
times a five sphere
\begin{equation}\label{eq:adsmetric}
\dd s^2 =-\frac{1}{R^2}\left( r^2 -\frac{r_{\rm H}^4}{r^2} \right) \dd t^2 +R^2 \,\frac{\dd r^2}{r^2 -\frac{r_{\rm H}^4}{r^2}} +\frac{r^2}{R^2} \,\dd\vec{x}^2 + R^2 \,\dd\Omega_5{}^2 ~.
\end{equation}
The \adsfive radius is related to the 't Hooft coupling of the dual gauge theory through $R^4 =g_{\rm YM}^2 N_c (\alpha')^2=\lambda (\alpha')^2$. The Hawking temperature is given by the horizon radius of the black hole $r_{\rm H}=\sqrt{\lambda}\alpha'\pi\,T$ and it is the temperature of the dual field theory.

As said in the introduction, there are two qualitatively different embeddings of the \mbox{D7-brane} in this AdS-Schwarzschild geometry. At large $m_q/(\sqrt{\lambda}\,T)$, with $m_q$ the quark mass, the tension of the D7-brane pulls enough to maintain itself outside of the black hole, ending at a finite value $r_0>r_{\rm H}$ of the radial coordinate. This is the so-called \textit{Minkowski} embedding. On the contrary, for small $m_q/(\sqrt{\lambda}\,T)$ the brane is forced to fall through the horizon thus inheriting the black hole structure. In this case one calls it the \textit{black hole} embedding. The conformal symmetry of the underlying AdS space can be used to set $r_{\rm H}$ to one. We will do this implicitly in the following by using the coordinate $z=r_{\rm H}/r$ and rescaling spacetime coordinates $\frac{r_{\rm H}}{R^2}(t,\vec{x})=T(t,\vec{x})\to (t,\vec{x})$. Aside from an overall $R^2$ factor, the line element is then
\be
\dd s^2 = \frac{1}{z^2} \left( -(1-z^4)\dd t^2 +\frac{\dd z^2}{1-z^4} +\dd\vec{x}^2 \right) +\dd\Omega_5^{\,2} ~,
\ee
which has the horizon sitting at $z =1$. We further write the $\rm S^5$ element as
\be
\dd\Omega_5^{\,2} = \dd\theta^2 +\sin^2\theta\,\dd\psi^2 +\cos^2\theta\,\dd\Omega_3^{\,2} ~,
\ee
where $\theta$ and $\psi$ parametrise the transverse directions to the brane.

The D7 wraps all of $\rm AdS_5$ (either down to the horizon or not) and the $\rm S^3$ inside the $\rm S^5$. Its action is the DBI action
\be
S_{\rm D7} =T_{\rm D7} \int\dd^8\xi \,\sqrt{-\det{\rm P}[G]} ~,
\ee
with ${\rm P}[G]$ the pullback of the bulk metric $G$ onto the brane. The profile of the embedding is characterised by its shape in the transverse coordinates. We will consider embeddings that have a simple profile characterised by the dependence of the S$^3$ radius on the \adsfive radial direction $(\theta(z),\psi=\rm const.)$.
Absorbing the tension and volume of the $\rm S^3$ into the normalisation, the action is
\be \label{eq:embedaction}
S_{\rm D7} = \int\dd z \,\frac{\cos^3\theta(z)}{z^5} \,\sqrt{1 +z^2(1-z^4)\,\theta'(z)^2\,{}} ~.
\ee

The embedding is obtained by solving the equation of motion for $\theta(z)$, which after some simplifications is
\bea\label{eq:embedequation}
0 &=& 3\sin\theta(z) \,\Big[-1+z^2\,(-1 +z^4) \,\theta'(z)^2 \,\Big] \\
&& +z\cos\theta(z) \Big[ (3+z^4)\,\theta'(z) +2z^2 \,(1-z^4)\,(2-z^4)\,\theta'(z)^3 +z(-1+z^4)\,\theta''(z) \Big] ~.\nn
\eea
For the Minkowski embedding one demands the brane to end outside the black hole, so the $\rm S^3$ has to shrink to zero size at $z_0$. Therefore one imposes $\theta(z_0)=\pi/2$. The second boundary condition comes from the requirement of a smooth ending without conical deficit. This imposes that $\theta'(z_0)\to\infty$ which we simulate in the numerical integration of \erf{eq:embedequation} by setting the derivative to $10^4$. In the black hole case, one sets the angle to some value \mbox{$\theta(1)=\theta_0$} at the horizon $z=1$, whereas the second boundary condition is $\theta'(1)=(3/4)\,\tan\theta_0$. This can be seen by demanding the embedding to be smooth at the horizon.\footnote{This boundary condition translates to Neumann boundary conditions when Fefferman--Graham coordinates are used as in \cite{Karch:2006bv}.} Figure \ref{fig:embeddings} shows the two embeddings for different values of the boundary conditions.
\FIGURE{\centering
\includegraphics[scale=1.1]{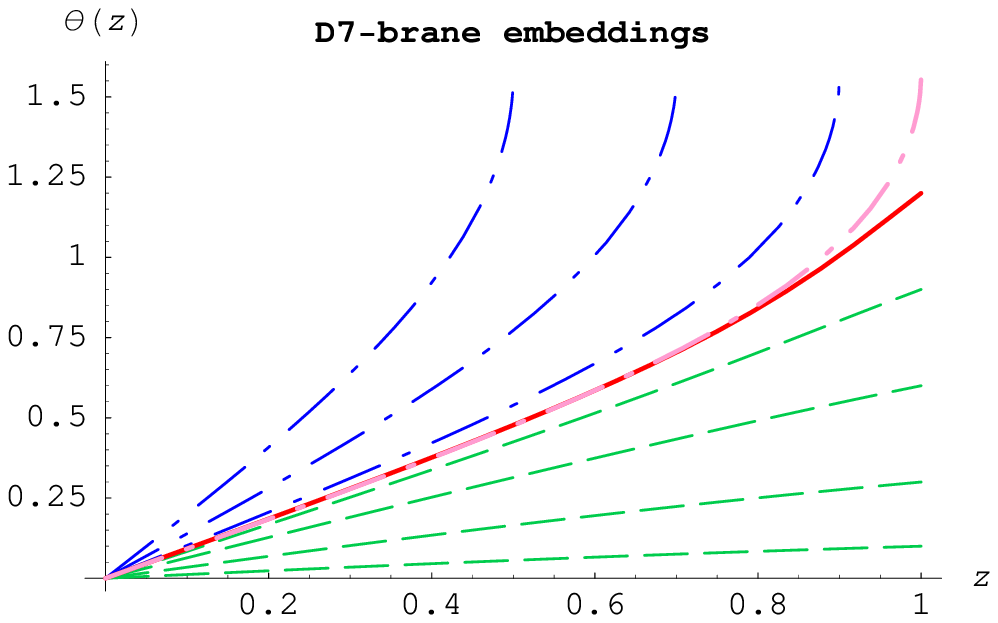}
\caption{\label{fig:embeddings}Minkowski (thin and thick dot-dashed) and black hole (dashed) D7-brane embeddings. We have also plotted the critical embedding (solid).}}

The solid curve in figure \ref{fig:embeddings} represents the critical embedding which is associated to the fundamental phase transition of the theory. In order to compute the critical embedding one needs to compute separately the free energy of the Minkowski and black hole embeddings which correspond to the \textit{same} quark mass, given by the derivative at the boundary. These free energies need to be renormalised,\footnote{One introduces an effective cutoff integrating down to $z=\epsilon\sim 0$, and later renormalises $\epsilon\to 0$.} where a holographic renormalization scheme may be used \cite{Henningson:1998gx, Balasubramanian:1999re, deHaro:2000xn}. For this particular setup this was done in \cite{Karch:2005ms}. The critical embedding corresponds to the configuration where the difference of free energies changes sign, with the free energy for the embedding given by the action \erf{eq:embedaction}, and with a discontinuity jump of the quark condensate.

The asymptotic expansion of the embedding
\be
\theta(z) =\Theta_0\,z + \Theta_2\,z^3 +\cO(z^5) +\ldots ~,
\ee
allows to obtain the quark mass
\be\label{eq.quarkmass}
m_q=\frac{1}{2} \,\Theta_0\sqrt{\lambda} \,T \equiv \Theta_0\,\Delta m(T) ~,
\ee
where we introduce the thermal rest mass factor $\Delta m(T)=\frac{1}{2}\sqrt{\lambda}\,T$, and also the condensate through \cite{Karch:2006bv}\footnote{We have used a definition of the mass that differs by a factor $\sqrt{2}$ from the one given in \cite{Karch:2006bv}.}
\be
\frac{\vev{q\bar q\,}}{\pi^2 T^2 \Delta m(T)} =\lim_{\epsilon\to 0} \,\frac{1}{\epsilon^3\,\sqrt{-\det{\rm P}[G]\Big|_{z=\epsilon}}} \,\frac{\delta S_{\rm reg.}}{\delta\,\theta(\epsilon)} =-4 \Theta_2 +\frac{2}{3}(\Theta_0)^3  ~.
\ee
This computation has been done in a variety of papers \cite{Babington:2003vm, Kirsch:2004km, Albash:2006ew}. The value of the quark mass for the critical embedding is $m_{\rm crit.} \approx 0.92 \,\Delta m(T)$. The physical situation corresponds to a discontinuous jump from a black hole embedding to a Minkowski embedding. It was also studied recently in \cite{Karch:2006bv} in the case of a curved boundary $\rm S^1\times\rm S^3$, where the deconfinement transition of the dual gauge theory was also analysed with similar results.

\section{Quasinormal Modes}  \label{sec:quasinormal}
In a black hole geometry there is a class of time-dependent fluctuations that behave as damped oscillations. They are called quasinormal modes to point out the difference with normal modes, that are non-damped oscillations. Quasinormal modes can be seen as excitations that dissipate their energy into the horizon or spreading it to infinity. In case we have an asymptotic AdS geometry all the energy is dissipated into the horizon, since the AdS curvature acts effectively as a box.

In the context of the AdS/CFT correspondence, a big black hole in AdS would correspond to a high-temperature plasma phase of the gauge theory. Properties of this plasma can be studied using correlation functions in the black hole background, and among other interesting results it is shown that the viscosity is non-zero, saturating a conjectured lower bound for the $\Nfour$ theory \cite{viscosity}. Therefore, thermal fluctuations of the plasma will be damped so it is natural to identify them with quasinormal modes of the black hole \cite{horowitz_hubeny, Cardoso:2001bb, Moss:2001ga, starinetsetal, Friess:2006kw}.

According to AdS/CFT, the black hole geometry corresponds to a strongly interacting quark-gluon plasma. From the point of view of mesons, there has been a deconfinement phase transition so they no longer provide a good description of fundamental degrees of freedom. Still, we could introduce a meson in the plasma and it will have a finite lifetime.

The usual prescription to identify the meson spectrum in the dual gravitational theory is to impose normalizable conditions at the boundary and regular conditions at the black hole horizon.  The spectrum turns out to be real and continous, so the associated two-point correlation functions will show a branch cut along the real axis pointing at the deconfinement of mesons. In \cite{starinetsetal} it was proposed to define the retarded Green function in terms of the quasinormal spectrum. Then, frequencies vanishing with the spatial momentum will correspond to wide and slow fluctuations thus describing the hydrodynamical properties of the plasma. Other quasinormal modes will also describe dissipation processes. In principle, quasinormal modes could describe unstable bound states of quarks in the plasma, with the same quantum numbers as the low temperature analog mesons and a finite lifetime. This will be true if distinct resonances appear in the spectral function, with a small enough width and an appropriate dispersion relation $\omega(\vec{q})$ so they can be interpreted as quasiparticles. In \cite{Kovtun:2005ev} it is shown that in $\Nfour$ plasma the quasinormal poles actually do not lead to resonances and that the high frequency behaviour is dominated by the underlying conformal symmetry. The case studied here is slightly different because for non-zero quark mass, the theory is already non-conformal and the structure of mesonic Green functions is given by an infinite set of discrete poles localized to the real axis, that we can identify with the stable spectrum of mesons. Also the fact that the plasma is strongly coupled for quarks and gluons does not affect mesons, because their interactions decrease at least as $\sim 1/N$ in the large $N$ limit, so from their point of view the plasma is a weakly coupled gas. This is confirmed by the zero drag force computed for mesons \cite{mesonmeltinHighspin}.

However, the relation between hypothetical bound states in the plasma and the actual mesons in the low temperature phase is not completely clear, since the wavefunction of a meson entering the plasma will probaly suffer strong non-linear effects before diluting. On the other hand, if we consider a single meson or a small number, the process of melting can be seen as a small fluctuation losing energy into the plasma and should be described by quasinormal modes.\footnote{Remember that we are considering low spin mesons only.}

In general, the problem of finding the quasinormal modes can be reduced to a quantum mechanical problem of scattering in a one-dimensional potential, see \cite{quasinormal_review} for reviews on quasinormal modes. In asymptotically flat spaces the potential vanishes both at infinity and at the horizon, so the solutions are in general a superposition of ingoing and outgoing waves. Quasinormal modes are defined imposing outgoing boundary conditions at infinity and infalling (ingoing) boundary conditions at the horizon. These conditions are quite restrictive and can be satisfied only by a discrete set of complex frequencies. In spaces that are asymptotically AdS the situation is similar at the horizon, but the potential does not vanish at infinity and a different set of boundary conditions should be considered. The usual choice is Dirichlet boundary conditions, although there could be other possibilities. In particular, for gravitational perturbations Neumann or Robin boundary conditions can also be considered \cite{Moss:2001ga}.

In the case that interests us, we introduce a set of D7 probes in an \adsfive black hole geometry. The fields living on the branes are gauge fields and two scalar fields $\theta$ and $\psi$, parameterising  the directions transverse to the brane. According to the correspondence, we can associate them to meson operators in the dual theory. From now on, we will be interested only in fluctuations of $\theta$, a scalar of mass squared $m^2=-3$, that in the dual theory maps to a meson operator of dimension $\Delta=3$, a quark bilinear. The normalizable modes of this field should correspond to scalar meson states in the dual theory.

From the two possible situations, let us consider D7-brane embeddings that fall into the black hole. Then, the induced metric on the branes has a horizon and there will be quasinormal modes in the spectrum of fluctuations of the brane. Since the black hole brane configuration corresponds to a deconfined situation for quarks, the meson spectrum associated to normal modes will be continuous \cite{Kruczenski:2003uq,Mateos:2006nu}.

For simplicity, we will consider only singlet states on ${\rm S}^3$, and since we are interested only in the mass and decay width, we will consider space-independent perturbations. However, it could be an interesting issue to see the effect of non-zero momentum with respect to the plasma rest frame on bound states. On general grounds, we expect that the states will be more stable.

Let us write the coordinate $\theta$ as $\theta(z,t) =\theta_0(z) +\vartheta(z,t)$, that is, embedding plus fluctuations over it. The action for the fluctuations is the DBI action of the brane
\begin{equation}
S_{\rm D7} \simeq \int \dd t\,\dd z \,\frac{\cos^3\left(\theta_0(z)+\vartheta\right)}{z^5}\sqrt{1-\frac{z^2}{1-z^4} (\partial_t\vartheta)^2+z^2(1-z^4) (\theta_0'(z) +\partial_z\vartheta)^2 } ~.
\end{equation}
If we consider small fluctuations, we can use the linearised equations of motion. Using $\vartheta(z,t) =\e^{-i\omega t} \vartheta(z)$, we are left with a second-order differential equation in $z$
\begin{equation} \label{eq:qnmodes}
\vartheta''(z) +A_1(z) \,\vartheta'(z) +(B(z)^2\,\omega^2+A_0(z)) \,\vartheta(z)=0 ~,
\end{equation}
where using
\begin{equation}
s(z) =1+ z^2(1-z^4) \,\theta_0'(z)^2 ~,
\end{equation}
the coefficients of the differential equation are
\begin{eqnarray}
B(z)   &=& \sqrt{s(z)}\,(1-z^4)^{-1} ~, \\
A_0(z) &=& s(z)\frac{3 \sec^2\theta_0(z)}{z^2 (1-z^4)} ~, \\
A_1(z) &=& -\frac{3+z^4}{z(1-z^4)} +6\theta_0'(z) \Big(\tan\theta_0(z) -z(2-z^4) \theta_0'(z) \Big) ~, \qquad
\end{eqnarray}
where we have used the equations of motion of the embedding (\ref{eq:embedequation}) to eliminate $\theta_0''(z)$.
Notice that if $\vartheta(z)$ is a solution for some frequency $\omega=\omega_\mathrm{R} -i \omega_\mathrm{I}$, then $\vartheta^*(z)$ is also a solution for a frequency $\omega'=-\omega^*=-\omega_\mathrm{R}-i\omega_\mathrm{I}$. This $\bZ_2$ symmetry allows to form real combinations of quasinormal modes, that will be true geometric deformations of the brane.

Close to the boundary of AdS ($z\to 0^+$), the differential equation is approximately
\begin{equation}
\vartheta''(z) -\frac{3}{z} \,\vartheta'(z) +\frac{3}{z^2} \,\vartheta(z) =0 ~,
\end{equation}
with solutions $\vartheta(z) \sim a\,z +b\,z^3$. According to the dictionary of the correspondence, we should set $a=0$ in order to study the spectrum of states. Otherwise, we will be introducing couplings for meson operators in the Lagrangian. Notice that the behaviour is universal for any embedding and frequency $\omega$.

Close to the horizon ($z\to 1^-$), the differential equation becomes
\begin{equation}
\vartheta''(z) +\frac{1}{z-1} \,\vartheta'(z) +\frac{\omega^2}{16(z-1)^2} \,\vartheta(z) =0 ~,
\end{equation}
with solutions $\vartheta(z)\sim a' (1-z)^{i\omega/4}+ b' (1-z)^{-i\omega/4}$. In this case, we will impose ingoing boundary conditions $a'=0$. The asymptotic behaviour is also universal for any embedding.

Equation \erf{eq:qnmodes} can be transformed into a Schr\"odinger equation, useful to derive analytic properties of the frequencies \cite{wkbmethods}. Using $\vartheta(z)=\sigma(z) f(z)$ with
\begin{equation}
\frac{\sigma'(z)}{\sigma(z)} =-\frac{1}{2}\left(A_1(z)+\frac{B'(z)}{B(z)}\right) ~,
\end{equation}
the equation for quasinormal modes is
\begin{equation} \label{eq:schrodinger}
\left( \frac{1}{B(z)} \frac{\dd}{\dd z} \left( \frac{1}{B(z)} \frac{\dd}{\dd z} \right) +\omega^2 -V(z) \right) f(z)=0~,
\end{equation}
so the Schr\"odinger equation is recovered after changing variables to the `tortoise' coordinate $\dd z_* =\sqrt{s(z)} \dd z/(1-z^4)$, that is defined such that the horizon is at $z_*\to \infty$. The potential is
\begin{equation}\label{eq:potential}
V(z) = -\frac{(1-z^4)^2}{s(z)} \left( A_0(z) +A_1(z) \frac{\sigma'(z)}{\sigma(z)} +\left(\frac{\sigma'(z)}{\sigma(z)}\right)^2 +\left(\frac{\sigma'(z)}{\sigma(z)}\right)' \,\right) ~.\qquad
\end{equation}

\begin{figure}[!htbp]
\centering
\includegraphics[scale=0.9]{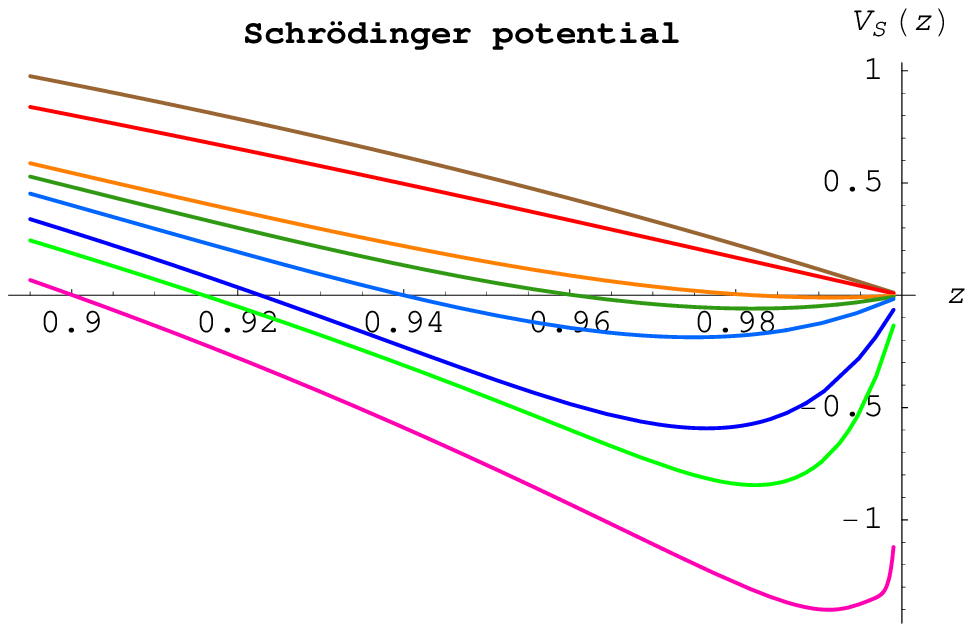}
\caption{\label{fig:potential}Schr\"odinger potential close to the horizon for different horizon embeddings, corresponding to $\theta_0(1)=0.1, 0.5, 0.83, 0.9, 1.0, 1.2, 1.3$ and $1.5$. As $\theta_0(1)$ increases, the potential develops a negative-valued well.}
\end{figure}

Consider now solutions that fall into the horizon, which in the $z_*$ coordinate are $f(z_*)=\exp(+i\omega z_*) \,\psi(z_*)$, multiply equation \erf{eq:schrodinger} by the conjugate solution $\psi^*(z_*)$ and integrate between the boundary and the horizon
\begin{equation}
\int_{z_*^b}^\infty \dd z_* \,( -\psi^*(z_*) \,\psi''(z_*) -2i\omega\,\psi^*(z_*) \,\psi'(z_*) +V(z_*) |\psi(z_*)|^2 \,) =0 ~.
\end{equation}
The solution $\psi(z_*)$ should vanish at the boundary and go to a constant at the horizon. The second-derivative term can be integrated by parts giving a result proportional to $|\psi'(z_*)|^2$. Taking the imaginary part of this equation and integrating by parts we find the relation
\begin{equation}
2 i \Im\omega \,\int_{z_*^b}^\infty \dd z_* \,\psi^*(z_*) \,\psi'(z_*) = -\omega^*|\psi(\infty)|^2 ~,
\end{equation}
that can be plugged back into the original equation
\begin{equation}
\int_{z_*^b}^\infty \dd z_* \,(\, |\psi'(z_*)|^2 +V(z_*) \,|\psi(z_*)|^2 \,) = -\frac{|\omega|^2 \,|\psi(\infty)|^2}{\Im\omega} ~.
\end{equation}
If the left hand side is positive, the imaginary part of the frequency must be negative. This depends on the value of the potential. The potential (\ref{eq:potential}) is non-negative between the boundary $z=0$ and the horizon $z=1$ for the massless case $\theta_0(z)=0$. For the massive case we found numerically that the potential is non-negative up to values $\theta_0(1)\sim 0.83$. Above this value the potential starts developing a well close to the horizon, so in principle there could be unstable modes of positive imaginary frequency.\footnote{We would like to thank R.~Myers, A.~Starinets and R.~Thomson for pointing out this fact to us.} The numerical plots of the potential are in figure \ref{fig:potential}.

Other qualitative properties of the frequencies can also be deduced from the shape of the potential. The frequencies of brane fluctuations are given by the energy spectrum of the \linebreak Schr\"odinger potential. For Minkowski embeddings, the potential is an infinite potential well, so the spectrum is real and discrete. For black hole embeddings, the potential is qualitatively the same close to the boundary, but it vanishes at the horizon. In both cases the potential develops a negative well close to the horizon as the embedding approaches the critical one separating Minkowski and black hole topologies. For Minkowski embeddings, the well starts developing for $z_0> 0.955$. If the well is deep enough, we expect that negative-energy bound states will appear. Bound states correspond to modes of tachyonic mass on Minkowski slices \cite{boundstates} and they are probably signalling an instability of the brane. We have checked the presence of tachyons up to $z_0=0.99$  with negative results. At $z_0=0.999$ we find that there is a single tachyon with $\omega \simeq 0.69\,i$, thus a true instability of the D-brane embedding. For black hole embeddings, the instability appears between $\theta_0(1)= 1.29$ and $\theta_0(1)=1.295$, where we find a mode with frequency $\omega\simeq 0.0014\,i$. Therefore, the first order transition occurs before these instabilities are present. Near-critical embeddings have been shown to be thermodynamically unstable \cite{Mateos:2007vn}, both instabilities seem to be related because the appearance of tachyonic modes coincide with the onset of the thermodynamical instability, in agreement with \cite{Buchel:2005nt}.

\subsection{Massless case} \label{ssec:massless}
We will consider first the simplest case of massless quarks, where the D7 embedding is trivial $\theta_0(z)=0$. The frequencies of the massless embedding can then be used as a starting point for the search of quasinormal frequencies in the massive case.

It is more convenient now to change our coordinate to $x=1-z^2$. The equation takes the simpler form
\begin{equation}
\vartheta''(x)+ \frac{1+(1-x)^2}{x(1-x)(2-x)} \,\vartheta'(x) +\left( \frac{\omega^2}{4 x^2 (1-x)(2-x)^2} +\frac{3}{4 x (1-x)^2 (2-x)} \right) \vartheta(x)=0 ~.
\end{equation}
This equation has four regular singularities $x=0,1,2,\infty$, with exponents $\{i\omega/4,-i\omega/4\}$, $\{1/2,3/2\}$, $\{w/4,-w/4\}$, $\{0,0\}$. Therefore, it is a Heun equation and we can follow the analysis described in \cite{Starinets:2002br} to compute the quasinormal frequencies.

Using the transformation
\begin{equation}
\vartheta(x) =x^{-i\omega/4}(x-1)^{3/2}(x-2)^{-\omega/4} \,y(x) ~,
\end{equation}
we can write the equation in the standard form for a Heun equation
\begin{equation}
y''(x) +\left( \frac{\gamma}{x} +\frac{\delta}{x-1} +\frac{\epsilon}{x-2}\right) y'(x) +\frac{\alpha\beta x-Q}{x(x-1)(x-2)} \,y(x) =0 ~,
\end{equation}
with parameters
\begin{equation}
\alpha=\beta=\frac{3}{2} -\frac{\omega}{4}(1+i) ~,\quad \gamma=1-i\frac{\omega}{2} ~,\quad \delta=2 ~,\quad \epsilon =1-\frac{\omega}{2} ~,
\end{equation}
and
\begin{equation}
Q = \frac{9}{4} -\frac{\omega}{4}(1+5 i) -\frac{\omega^2}{8}(2-i) ~.
\end{equation}
Quasinormal modes correspond to solutions of this equation defined in the interval $[0,1]$ with boundary conditions $y(0)=y(1)=1$, that select the appropriate normalizable and ingoing behaviour of the solution.

We can find local solutions using the Frobenius method close to the singularities. Usually, a solution with boundary condition $y(0)=1$ will be a superposition of solutions with exponents $1/2$ and $3/2$ close to the AdS boundary ($x=1$). The boundary condition $y(1)=1$ can be satisfied only for a discrete set of frequencies $\omega$. These values can be computed imposing matching conditions at some intermediate point for the Frobenius series although we will need a large number of terms and the convergence gets worse for higher frequencies. There is an alternative method based on the improved convergence of the solutions. Normal solutions are convergent for $|x|<1$, but for some values of the parameters the solutions can converge for $|x|<2$. This happens when we have two possible solutions for the recursion relations of the Frobenius series at the horizon, so we can satisfy the requirements for quasinormal modes at the AdS boundary. The condition of convergence boils down to a transcendental equation for the frequencies in the form of a continued fraction (see \cite{Leaver,Starinets:2002br} for details).

The coefficients of the Frobenius series at $x=0$ should satisfy the recursion relation
\begin{equation}
a_{n+2} +A_n(\omega) \,a_{n+1} +B_n(\omega) \,a_n=0 ~, \quad n\ge 2 ~,
\end{equation}
where
\bea
A_n(\omega) &=& -\frac{(n+1)(2\delta+\epsilon+3(n+\gamma))+Q}{2(n+2)(n+1+\gamma)} ~, \\
B_n(\omega) &=& \frac{(n+\alpha)(n+\beta)}{2(n+2)(n+1+\gamma)} ~,
\eea
and $a_0=1, ~a_1=Q/2\gamma$. Then, using the recursive definition
\begin{equation}
r_n =\frac{a_{n+1}}{a_n} = -\frac{B_n(\omega)}{A_n(\omega)+r_{n+1}} ~,
\end{equation}
the convergence condition is
\begin{equation}
r_0 =\frac{Q}{2\gamma} ~.
\end{equation}
Using this formula, we can easily compute numerically the quasinormal frequencies with high precision. In order to do that, we cut the fraction at some large value of $n$ and use the asymptotic value $r_n=1/2-(2+\omega)/4n$. The frequencies for the first ten modes are compiled in table \ref{table1}. Higher modes $k\gg 1$ seem to have the asymptotic behaviour
\begin{equation}
\omega_k\simeq 1.219-0.779 i+(2k-1)(1-i) ~.
\end{equation}
\begin{table}[!htbp]
\hspace*{4ex}
\begin{tabular}{cc}
\parbox[c]{.35\textwidth}{%
\vspace*{-16em}
\begin{tabular}{|c|c|c|} \hline
$k$ & $\Re(\omega_k)$ & $\Im(\omega_k)$ \\ \hline\hline
1  & \hfill 2.1988  & \hfill $-1.7595$ \\
2  & \hfill 4.2119  & \hfill $-3.7749$ \\
3  & \hfill 6.2155  & \hfill $-5.7773$ \\
4  & \hfill 8.2172  & \hfill $-7.7781$ \\
5  & 10.2181        & \hfill $-9.7785$  \\
6  & 12.2186        & $-11.7787$ \\
7  & 14.2180        & $-13.7788$ \\
8  & 16.2193        & $-15.7789$ \\
9  & 18.2195        & $-17.7790$ \\
10 & 20.2183        & $-19.7790$ \\ \hline
\end{tabular}} &
\parbox[t]{.6\textwidth}{\includegraphics[scale=0.87]{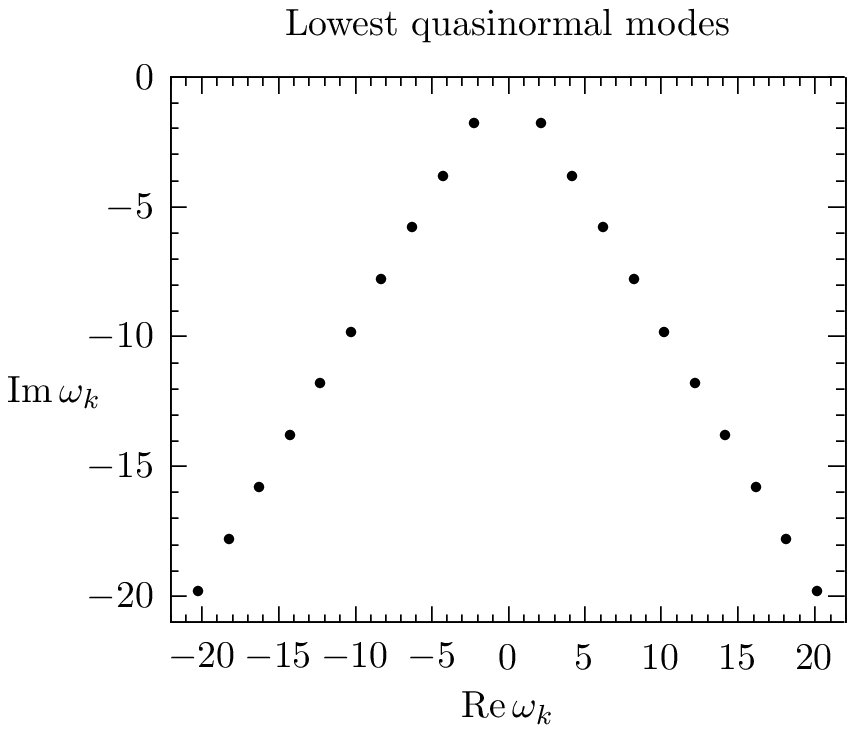}}
\end{tabular}
\caption{\label{table1} Quasinormal frequencies for massless quarks using the convergence method ($n=200$ for $k<10$ and $n=250$ for $k=10$).}
\end{table}

\subsection{Massive case} \label{sec:massive}
When we consider massive quarks, associated to a non-trivial D7 profile $\theta_0(1)\neq 0$, the situation gets much more involved. First of all, we only know the embedding numerically, and second, we know that it is not an analytic function, since its expansion close to AdS boundary involves logarithmic terms. So we have to use  mainly a numerical approach to compute the quasinormal frequencies.

We use a shooting method to compute the frequencies. We approximate our numerical embeddings by a power expansion in order to keep computation time bounded. This is our main source of error, typically we trust our results up to the fourth decimal. Close to the singular points we approximate the modes by a series expansion with the right behaviour, as $z^3$ at the AdS boundary ($z=0$) and like $(1-z)^{-i\omega/4}$ at the horizon ($z=1$). We use then the series expansion at the horizon to give initial values for the numerical solution. Close to the boundary, we try to match smoothly the numerical solution with the series expansion at $z=0$. The matching points between the numerical solution and the series expansions are selected so that the error in the differential equation coming from the series is less than $10^{-8}$, all values lying in the intervals $[0.0375,0.15]$ and $[0.85,0.9625]$.

The matching can be done only for discrete values of the frequency that we find exploring the complex $\omega$ plane. The frequencies that we have found for the massless case are very useful to give a starting point for our search, and we also use them to check the method for the massless case. Since the method becomes quite expensive in terms of CPU time for higher modes, especially when we approach the critical embedding $\theta_0(1)=\pi/2$, we have limited to the first three modes. The results can be found in tables \ref{table2}, \ref{table3} and \ref{table4} and in figure \ref{fig:freqs}. We see that the evolution drives the quasinormal frequencies from their values in the massless embedding toward values closer to the real axis. Unfortunately, we are not able to reach the limiting embedding numerically, so we cannot confirm what is the endpoint of the evolution. It would be interesting to make an improved numerical analysis or an analytic computation of quasinormal frequencies for near-limiting embeddings to address this issue.

\FIGURE{\centering
\includegraphics{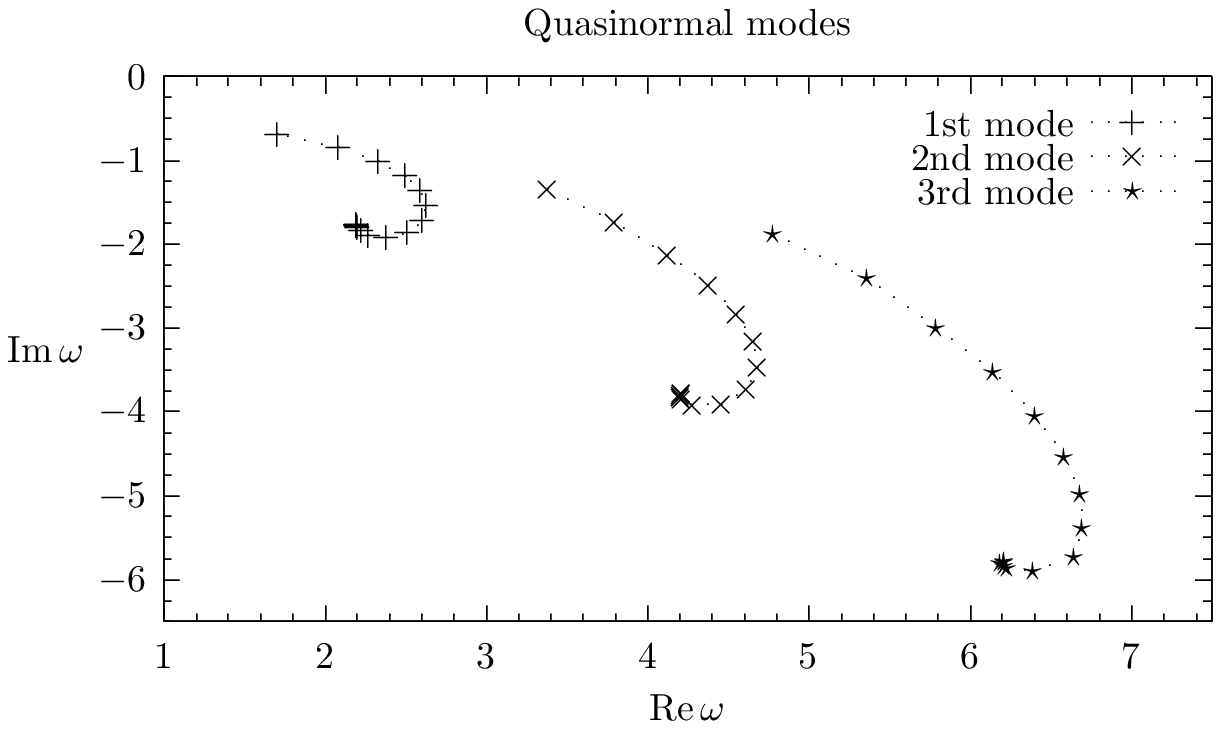}
\caption{\label{fig:freqs} Evolution of the first three quasinormal frequencies in the complex plane as we change the embedding. All of them show the same behaviour when $\theta_0$ is increased: firstly they go to the right to reach a returning point and start moving to the left decreasing their imaginary part.}}

We have checked the results using a purely numerical approach, the relaxation method, where the differential equation is converted into a finite difference equation. In this method we use the numerical values of the embedding, lowering significantly this source of error. We find that the results for the first mode have a very good agreement, typically the absolute relative error in the frequencies is of order $10^{-4}$. The details of the relaxation method are discussed in the appendix.

\DOUBLETABLE{%
\begin{tabular}{|c|c|c|} \hline
$\theta_0(1)$ & $\Re(\omega_1)$ & $\Im(\omega_1)$ \\ \hline\hline
0.00 & 2.1988 & $-1.7595$ \\
0.10 & 2.1989 & $-1.7636$ \\
0.20 & 2.1999 & $-1.7765$ \\
0.25 & 2.2016 & $-1.7868$ \\
0.30 & 2.2051 & $-1.8001$ \\
0.40 & 2.2225 & $-1.8371$ \\
0.50 & 2.2712 & $-1.8855$ \\
0.60 & 2.3785 & $-1.9133$ \\
0.70 & 2.5153 & $-1.8546$ \\
0.80 & 2.6057 & $-1.7172$ \\
0.90 & 2.6299 & $-1.5454$ \\
1.00 & 2.5935 & $-1.3665$ \\
1.10 & 2.4972 & $-1.1867$ \\
1.20 & 2.3342 & $-1.0135$ \\
1.30 & 2.0866 & $-0.8488$ \\
1.40 & 1.7078 & $-0.6846$ \\ \hline
\end{tabular}}{%
\begin{tabular}{|c|c|c|} \hline
$\theta_0(1)$ & $\Re(\omega_2)$ & $\Im(\omega_2)$ \\ \hline\hline
0.00 & 4.2119 & $-3.7749$ \\
0.10 & 4.2090 & $-3.7808$ \\
0.20 & 4.2035 & $-3.8020$ \\
0.25 & 4.2057 & $-3.8218$ \\
0.30 & 4.2136 & $-3.8477$ \\
0.40 & 4.2818 & $-3.9122$ \\
0.50 & 4.4607 & $-3.9061$ \\
0.60 & 4.6172 & $-3.7287$ \\
0.70 & 4.6822 & $-3.4663$ \\
0.80 & 4.6585 & $-3.1643$ \\
0.90 & 4.5558 & $-2.8370$ \\
1.00 & 4.3784 & $-2.4866$ \\
1.10 & 4.1243 & $-2.1242$ \\
1.20 & 3.7945 & $-1.7345$ \\
1.30 & 3.3796 & $-1.3415$ \\ \hline
\end{tabular}}{\label{table2}First quasinormal frequencies for different D7 embeddings.}{\label{table3}Second quasinormal frequencies for different D7 embeddings.}

\begin{table}[!htbp]
\bcen
\begin{tabular}{|c|c|c|} \hline
$\theta_0(1)$ & $\Re(\omega_3)$ & $\Im(\omega_3)$ \\ \hline\hline
0.00 & 6.2155 & $-5.7773$ \\
0.10 & 6.2064 & $-5.7813$ \\
0.20 & 6.1851 & $-5.8052$ \\
0.25 & 6.2067 & $-5.8341$ \\
0.30 & 6.2299 & $-5.8666$ \\
0.40 & 6.3867 & $-5.8924$ \\
0.50 & 6.6466 & $-5.7291$ \\
0.60 & 6.6925 & $-5.3760$ \\
0.70 & 6.6829 & $-4.9753$ \\
0.80 & 6.5846 & $-4.5276$ \\
0.90 & 6.4048 & $-4.0423$ \\
1.00 & 6.1417 & $-3.5153$ \\
1.10 & 5.7917 & $-2.9881$ \\
1.20 & 5.3629 & $-2.4019$ \\
1.30 & 4.7783 & $-1.8763$ \\ \hline
\end{tabular}
\ecen
\caption{\label{table4}Third quasinormal frequencies for different D7 embeddings.}
\end{table}

\section{Meson masses and lifetimes}  \label{sec:mesonmasses}

We have identified small fluctuations of D7-probe branes with the low energy spectrum of mesons. In the zero temperature case ($T=0$), the spectrum of mesons is given by regular and normalizable modes on the brane. The bare mass of the quarks $m_q$ is identified with the asymptotic properties of the embedding, as in \erf{eq.quarkmass}. If the mass is finite, then the embedding ends at a finite value of the radial coordinate, providing an IR cutoff for the modes on the brane. The spectrum is discrete with a mass gap $M\sim m_q/\sqrt{\lambda}$ and grows linearly \cite{Karch:2002sh,Kruczenski:2003be}. If the bare mass is zero, then the induced metric on the brane is conformal ${\rm AdS}_5\times {\rm S}^3$ and the spectrum becomes continuous.

For Minkowski embeddings, the spectrum will be similar to the zero temperature case for $m_q/\Delta m(T) \gg 1$. We show the first modes of the meson spectrum for several embeddings in table \ref{tableMass} and figure \ref{fig:massNH}, notice that the mass gap grows linearly with the quark mass for embeddings with $m_q/\Delta m(T) > 1$.
\FIGURE{\centering
\includegraphics[scale=1]{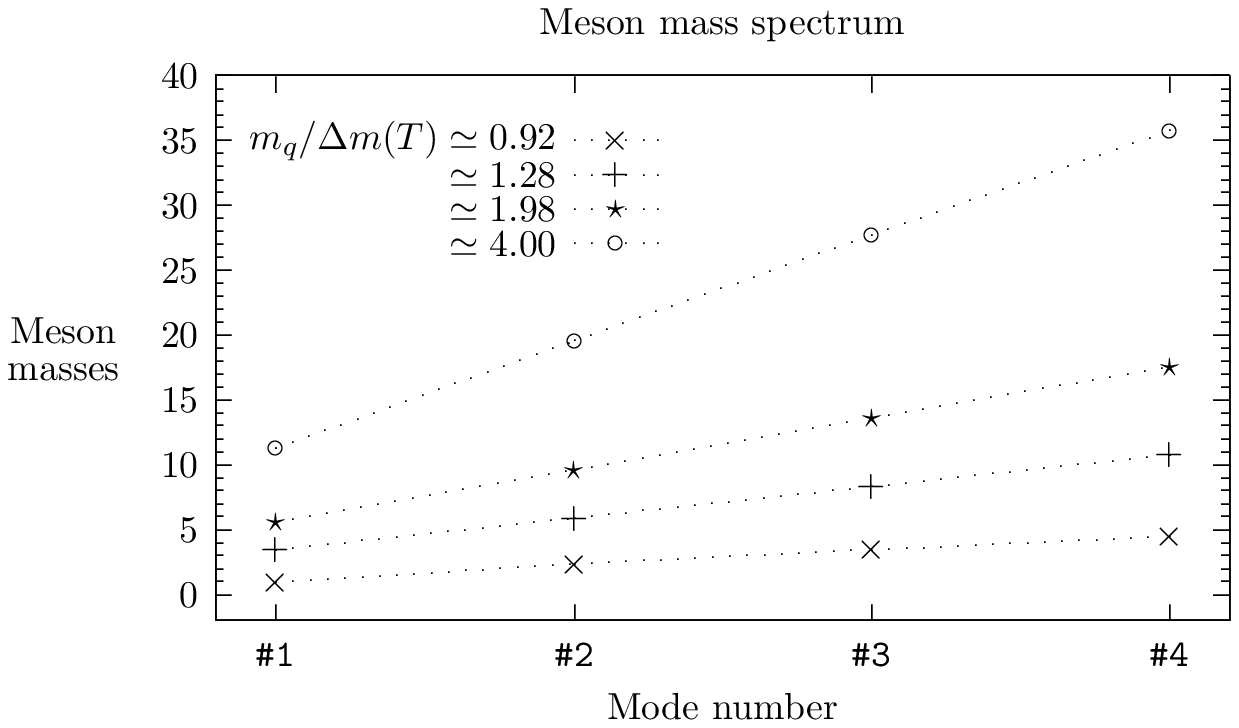}
\caption{\label{fig:massNH} First modes of the meson spectrum for Minkowski embeddings ending at $z_0=0.25,\, 0.5,\, 0.75,\, 0.99$. We observe that the spectrum grows linearly and the mass gap increases with the distance of the embedding to the horizon $z_+=1$.}}

\begin{table}[!htbp]
\centering
\begin{tabular}{|c|c|c|c|c|} \hline
$m_q/\Delta m(T)$ & \texttt{1} & \texttt{2} & \texttt{3} & \texttt{4}\\ \hline\hline
0.91784 & ~0.9589 & ~2.3411 & ~3.4388 & ~\,4.5087 \\
1.27999 & ~3.4357 & ~5.9163 & ~8.3647 & 10.7981 \\
1.98432 & ~5.5717 & ~9.6391 & 13.6312 & 17.5994 \\
3.99805 & 11.3081 & 19.5905 & 27.7128 & 35.7863 \\ \hline
\end{tabular}
\caption{\label{tableMass}First meson masses in temperature units for different quark masses in Minkowski embeddings.}
\end{table}

Below the critical mass $m_q \simeq 0.92 \,\Delta m(T)$, there can be a first or a second order transition to a black hole embedding \cite{Babington:2003vm, Karch:2006bv, Albash:2006ew,Kirsch:2004km, Ghoroku:2005tf, Mateos:2006nu, Frolov:2006tc}. Free energy arguments show that the branch of black hole embeddings reached by the first order transition will dominate. However, the second order transition is interesting because its properties are similar to type II critical collapse of black holes and black hole/black string merger transitions \cite{Mateos:2006nu, Frolov:2006tc}. In black hole embeddings, the theory is in a deconfined phase and the spectrum of mesons is continuous. From the geometric point of view, this is due to the presence of a horizon.

\begin{figure}[!htbp]
\includegraphics{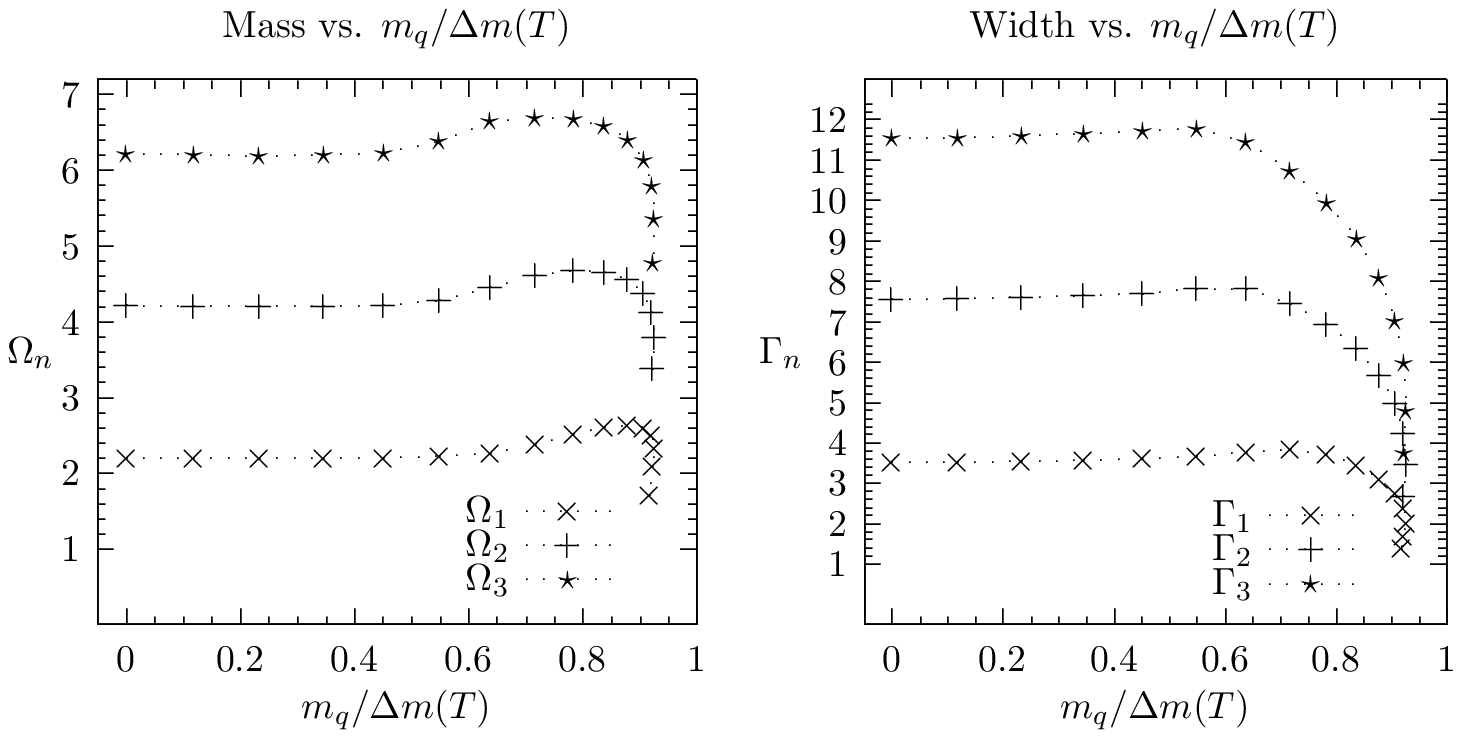} 
\caption{\label{fig:masswidth} (Left) Quasinormal masses in temperature units for the first three modes as a function of the bare quark masses, and (right) quasinormal widths in temperature units for the first three modes as a function of the bare quark masses.}
\end{figure}

Let us consider the results of the previous section. The mass and the width of quasinormal modes are roughly proportional, and proportional to the temperature (see Fig.~\ref{fig:masswidth}). As we increase the quark mass there is some change, close to the critical value where the two branches of black hole embeddings merge. The width decreases appreciably, specially for higher modes, while the mass does not change as much. In any case, before the first order phase transition the mass and the width are of the same order, which makes a quasiparticle interpretation unlikely.

The melting is therefore characterized by the temperature, showing little dependence on the quark mass {\em after the transition}. This implies that mesons made of light quarks will start melting at lower temperatures but will have longer lifetimes just after the transition. Considering mesons made of quarks of fixed mass, heavier mesons with the same quantum numbers (in our case, zero spin mesons associated to the scalar operator $\overline{q} q$) will initially decay through higher quasinormal modes, so they will lose energy faster. Then, a heavy meson that goes through a plasma region will usually emerge as a lower mass state. If the meson does not emerge, then the decay is dominated by the lowest quasinormal mode at large times and becomes universal, it is no longer possible to distinguish the original state. This means that any scattering process will in principle increase the energy and entropy of the plasma, as we expect from a dissipative medium.

\section{Conclusions and Outlook}  \label{sec:conclude}

We have proposed a holographic picture (of the late stages)  of the melting process of low spin mesons in the quark-gluon plasma. The important ingredient are D7-brane embeddings in AdS black hole backgrounds that do reach down to the horizon. The induced metric on the world volume of these D7-branes is itself a black hole and therefore it makes sense to compute the quasinormal modes of the brane fluctuations. These modes describe the dissipation of the energy of mesonic excitations by the plasma, in particular the melting of mesons. One important point is that this process of melting in the holographic plasma is only available for mesons built out of quarks with masses up to $m_q = 0.92\,\sqrt{\lambda}\,T/2$, where $T$ is the plasma temperature and $\lambda$ is the 't Hooft coupling. Heavier quarks are represented by D7-branes with no horizon on their world volume and therefore have stable meson excitations.

The melting of quarkonium states is of high importance in the physics of the quark-gluon plasma. It has long been regarded to be one of the cleanest signatures of plasma formation. In particular, quarkonium states such as the $J/\psi$ meson are expected to melt in the quark-gluon plasma and therefore the abundance of these particles measured in processes where quark-gluon plasma formation takes place should drop significantly if compared to nuclear collisions without plasma formation.

Although the model we have considered here is quite far from QCD it is still interesting to have a look to this problem from our perspective of holographic meson melting. We found that  the melting process takes place only for quarkonium states built out of quarks with masses of at most the order of the temperature of the plasma. The mass of the charm quark is $m_c \approx 1.4$ GeV and the RHIC temperatures is $T_\mathrm{RHIC}\lessapprox 300$ MeV. AdS predicts critical mass to temperature ratios of $1-2$ if we use the recent estimates on how to relate $5.5<\lambda<6\pi$ to QCD \cite{Gubser:2006qh}. It is quite interesting that very recently meson melting has been considered in a real time approach in Hard Thermal Loop resummed perturbation theory \cite{Laine:2006ns}. There the authors found an imaginary part in the static quark anti-quark potential giving rise to a decay width and that this decay width can be ignored for quarks heavier that $m_q = 12\pi T/g^2$.  One is tempted to speculate that this is the weak coupling QCD analog of the holographic value $m_q=0.92 \sqrt{\lambda} T/2$ and that there exists an interpolating function $f(\lambda)$ such that $m_q = f(\lambda) T$, where $f(\lambda) \approx \sqrt{\lambda}$ at strong coupling and $f(\lambda) \approx 1/\lambda$ at weak coupling.

The biggest drawback of the AdS-model is that the background corresponds to plasmas made up only of particles in the adjoint representation. Although in a background with dynamical quarks included one still expects heavy quarks being reasonably well modelled by D-brane embeddings the presence of fundamental quarks in the deconfined plasma might change the dissociation rates even for heavy quarks in a drastic way. In any case, the quasinormal modes on D-branes embedded in gravity duals of gauge theories offer a unique way of studying quarkonium dissociation in a holographic way.

It is of high interest to apply our approach to meson melting also to other models, such as models with non-Abelian chiral symmetries \cite{Sakai:2004cn} or the phenomenological holographic models of QCD developed in \cite{holoQCD}. In view of the above mentioned problem of $J/\psi$ suppression it would be of extreme interest to have a phenomenological holographic QCD model that includes heavy flavours such as the charm quark.

In this paper we have only considered modes with vanishing momentum on the S$^3$ as well as on $\bbR^3$. Especially, the dependence of the quasinormal modes on momentum relative to the rest frame of the plasma should be quite interesting. Furthermore one could study the modes of the other fields on the D7-brane world volume, such as the vector fields whose excitations correspond to vector mesons. We also have found evidence that quasinormal modes show an interesting behaviour when the system goes through the second order phase transition, allowing a continuous connection of both phases. We plan to come back to these questions in future publications \cite{us.future}.

\acknowledgments
The research of K.\,L. is supported by the Ministerio de Ciencia y Tecnolog\'{\i}a through a Ram\'on y Cajal contract and by the Plan Nacional de Altas Energ\'{\i}as FPA-2006-05485. The research of S.\,M. is supported by an FPI 01/0728/2004 grant from Comunidad de Madrid and by the Plan Nacional de Altas Energ\'{\i}as FPA-2006-05485. S.\,M. also wants to thank G. S\'anchez for her support.
We would like to thank Gert Aarts, Adi Armoni, Ioannis Bakas, Jos\'e Barb\'on, Gaetano Bertoldi, Margarita Garc\'{\i}a P\'erez, Prem Kumar, Esperanza L\'opez,  Cristina Manuel, Wolfgang M\"uck, Asad Naqvi and Carlos N\'u\~nez for useful discussion and their interest in our work.
We would also like to thank R.~Myers, A.~Starinets and R.~Thomson for useful comments and for pointing out to us some mistakes in the discussion of the Schr\"odinger potential in an earlier version.

\appendix

\section{Relaxation Method}  \label{app:relaxmethod}
It is common practice to use the shooting method to solve two-point boundary value differential equations. Indeed, we have used this method to compute the first three quasinormal modes of the massive embedding for a variety of bare quark masses at the horizon. However, in order to check at least those results for the first mode, we follow the spirit of \textit{``always shoot first and only then relax".} The reference for this appendix is Numerical Recipes \cite{nrinc}.

The idea behind this method consists in replacing the differential equations by a finite-difference system (FDE) on a grid. One will then modify the value of the dependent variables at each point \textit{relaxing} to the configuration which solves the differential equation. In our case, we will convert our second order complex ODE into a set of four first order equations, with the real and imaginary parts separated, and then into an FDE.

In general one has the set of $N$ ODE's
\be \label{eq:odesys}
\frac{\dd y_i(x)}{\dd x} =g_i(x,y_1,\ldots,y_N;\lambda) ~,
\ee
where each dependent variable depends on the others and itself, on the independent variable $x$, and possibly on additional parameters, like $\lambda$ above. In our case this (complex) parameter takes the role of the quasinormal frequency. Those extra parameters can be embedded in the problem giving equations for them too
\bea
\left\{ \ba{rcl} y_{N+1} &\equiv& \lambda~,\\[-1em] \\ \displaystyle \frac{\dd y_{N+1}}{\dd x} &=& 0~, ~~\hbox{since it is constant~.}\ea \right.
\eea
The solution to the problem involves $N\times M$ values, for the $N$ dependent variables in a grid of $M$ points. Concerning boundary conditions, for the system to be determined one needs $N$ of them, supplying it with extra boundary conditions for the extra parameters if present.

The system is discretized as usual
\be
x \to \frac{1}{2}(x_k +x_{k-1}) ~, ~~ y \to \frac{1}{2}(y_k +y_{k-1}) ~,
\ee
for points in the bulk (not in the boundaries). One may arrange the whole set of $y_i$'s in a column vector ${\mathbf y}_k =(y_1,\ldots,y_N,y_{N+1})_k{}^T$, where the subscript $k$ refers to evaluation at the point $x_k,~k=1,\ldots,M.$ With this matrix notation, the system \erf{eq:odesys} is rewritten as
\be
0 = \bE_k \equiv \by_k -\by_{k-1} -(x_k-x_{k-1}) \,\bg_k(x_k,x_{k-1},\by_k,\by_{k-1}) ~,~~k=2,\ldots,M ~,
\ee
where the $\bE_k$ are the aforementioned FDE's. These are the equations that we need to fulfil. Notice there are $N$ equations at $M-1$ points, so the remaining $N$ equations are just given by the boundary conditions. We will set $n_1$ of them on the left at $x_1$, called $\bE_1$, and the rest $n_2=N-n_1$ at $x_M$, called $\bE_{M+1}$ (no typo in the subscript!).

Now one is set with all the necessary material. Suppose one has a good ansatz that nearly solves the FDE's $\bE_k$. By shifting each solution $\by_k\to\by_k+\Delta\by_k$ and Taylor expanding in the shift, one obtains a relation
\bea
0 =\bE_k(\by+\Delta\by) &\simeq& \bE_k(\by_k,\by_{k-1}) +\sum_{n=1}^N \frac{\partial\bE_k}{\partial y_{n,k-1}} \,\Delta y_{n,k-1} +\sum_{n=1}^N \frac{\partial\bE_k}{\partial y_{n,k}} \,\Delta y_{n,k}~,~~ \\
\then ~-E_{j,k} &=& \sum_{n=1}^N \Big( S_{j,n} \,\Delta y_{n,k-1} \Big) +\sum_{n=N+1}^{2N} \Big( S_{j,n} \,\Delta y_{n-N,k} \Big) ~,
\eea
which by inverting it allows to find the $\Delta\by_k$ that improve the solution. This is done by merging the two differentials
\be
S_{j,n} =\frac{\partial E_{j,k}}{\partial y_{n,k-1}} ~,~~ S_{j,n+N} =\frac{\partial E_{j,k}}{\partial y_{n,k}} ~,~~~ n=1,\ldots,N ~,
\ee
in a $N\times 2N$ matrix, for each bulk grid position $x_k$. For the boundaries the expressions follow equally
\bea
-E_{j,1} &=& \sum_{n=1}^N S_{j,n} \,\Delta y_{n,1} =\sum_{n=1}^N \frac{\partial E_{j,1}}{\partial y_{n,1}} \,\Delta y_{n,1} ~, \quad j_2+1,\ldots,N ~, \\
-E_{j,M+1} &=& \sum_{n=1}^N S_{j,n} \,\Delta y_{n,M} =\sum_{n=1}^N \frac{\partial E_{j,M+1}}{\partial y_{n,M}} \,\Delta y_{n,M} ~, \quad j=1,\ldots,n_2 ~,
\eea
where $n$ runs in both from $1$ to $N$. The whole $(NM\times NM)$ $S$ matrix possess a block diagonal structure. This allows for a somewhat fast Gaussian elimination, since off-diagonal entries are already zero, and forms the basis of an iterative process which can be put to run until one reaches a desired accuracy in the result. Concerning this last point, we computed the error of our solution as
\be
\mathtt{err} = \frac{1}{MN} \sum_{k=1}^M \sum_{j=1}^N \left| \frac{\Delta y[j][k]}{\mathtt{scalevar}[j]} \right| < \mathtt{conv}~,
\ee
where $\mathtt{scalevar}[j]$ is an associated scale for each of the dependent variables (e.g. the value at the midpoint or so). The idea is that when that averaged value of the shift to get a better solution is smaller than \texttt{conv}, we accept the former values we had as the actual solution. In our computations we set $\mathtt{conv}=10^{-6}$.


\end{document}